\documentclass[journal]{IEEEtran}
\usepackage{authblk}
\usepackage{cite}
\usepackage{amsmath}
\usepackage{amsthm}
\usepackage{amssymb}
\usepackage{graphicx}
\usepackage{subcaption}
\usepackage{url}
\usepackage{epstopdf}
\usepackage{placeins}
\usepackage{epsfig}
\usepackage{amsfonts}
\usepackage{balance}
\usepackage[ruled,vlined]{algorithm2e}
\newtheorem{theorem}{Theorem}

\newtheorem{lemma}{Lemma}
\newtheorem{example}{Example}
\newtheorem{corollary}{Corollary}
\newcommand{\mytilde}{\raise.17ex\hbox{$\scriptstyle\mathtt{‌​\sim}$}}
\begin{document}
\title{Index Coding in Vehicle to Vehicle Communication}

\author{Jesy Pachat, \IEEEmembership{~Member,~IEEE},  Nujoom Sageer Karat,  \IEEEmembership{~Student Member,~IEEE}, Deepthi P. P., \IEEEmembership{~Member,~IEEE}, and B. Sundar Rajan, \IEEEmembership{Fellow,~IEEE}
\thanks{Copyright (c) 2015 IEEE. Personal use of this material is permitted. However, permission to use this material for any other purposes must be obtained from the IEEE by sending a request to pubs-permissions@ieee.org.}
\thanks{Jesy Pachat and Deepthi P. P. are with the Electronics and Communication Department, National Institute of Technology, Calicut, 673601, India (e-mail:jesypachat@gmail.com, deepthi@nitc.ac.in).}
\thanks{N. S. Karat and B. S. Rajan are with the Department of Electrical Communication Engineering, Indian Institute of Science, Bangalore, 560012, India (e-mail: nujoom@iisc.ac.in, bsrajan@iisc.ac.in).}
}%

\markboth{IEEE Transactions on Vehicular Technology,~Vol.~XX, No.~XX, XXX~2020}
{}

\maketitle

\begin{abstract} 
Vehicle to Vehicle (V2V) communication phase is an integral part of collaborative message dissemination in vehicular ad-hoc networks (VANETs). In this work, we apply index coding techniques to reduce the number of transmissions required for data exchange. 
The index coding problem has a sender, which tries to meet the demands of several receivers in a minimum number of transmissions.  All these receivers have some prior knowledge of the messages, known as the side-information. In this work, we consider a particular case of the index coding problem, where multiple nodes want to share information among them. Under this set up, lower bound on the number of transmissions is established when the cardinality of side-information is the same. An optimal solution to achieve the bound in a special case of VANET scenario is presented. For this special case, we consider the link between the nodes to be error-prone, and in this setting, we construct optimal linear error correcting index codes. 
\end{abstract}


\IEEEpeerreviewmaketitle

%%%%%%%%%%%%%%%%%%%%%%%%%%%%%%%%%%%%%%%%%%%%%%%%%%%%%%%%%%%%%%%%%%%%%%%%%
\section{INTRODUCTION}
Vehicular Adhoc Networks (VANETs) have gained interest with their important roles in intelligent transport systems. The implementation of intelligent transportation systems requires the coordination of different entities and support for different modes of communication \cite{sur}. The early steps towards the Vehicle-to-Everything (V2X) communication specifications are built on group communications and proximity service features, both initially developed for mission-critical communication. These efforts strengthened the design of a new standard called cellular V2X (CV2X) supported by 5G systems. The specification is an extension of LTE, and it has been released by the 3GPP in Release 14 \cite{3gppRel14}. This standard incorporates C-V2X mode 4, which is explicitly designed for V2V communications using the PC5 sidelink interface without any support for cellular infrastructure \cite{v2xAna}, \cite{veh_mag}.The coexistence over the unlicensed spectrum of VANET users and cellular V2X users has been studied in \cite{coexist}.

The primary motivation for the vehicle to vehicle communication is to enable safety as well as infotainment services to the users. For this to be a reality, the content distribution among the vehicles, and between the infrastructure and the vehicles are necessary. For VANETs, direct data transfer to vehicles typically occur via cellular network solutions or solutions based on dedicated short-range communications (DSRC) \cite{dsrc}. Despite the efficiency of cellular network-based solutions, it demands excellent coverage for a large number of users, and the cost associated with network usage is high. The limited roadside coverage of the DSRC solutions is also challenging. This motivates the use of  cooperative data dissemination in VANETs.

Collaborative data dissemination techniques in VANETs are well discussed in literature. These techniques mainly consider network coding \cite{NC} aspects for reducing the download time and increasing the throughput. The problem of collaborative content distribution in vehicular ad-hoc network (VANET) is addressed in \cite{ZLS, col1, col2, col3, col4}. This paper looks at the context of index coding (IC) \cite{BiK} for the distribution of messages from Road Side Unit (RSU) among a group of vehicles while using cooperative content distribution techniques. 

 In this work, we view the V2V phase as simultaneous index coding problems, which is applicable in device-to-device (D2D) communication problem \cite{D2D1, JCM}. We prove that there is a considerable improvement in the download completion time by using index coding. For a special case, we provide an index coding scheme which is optimal. Furthermore, the cooperative data exchange algorithm in \cite{RSS} is extended to the V2V data exchange phase and the improvement is characterized. 

Whereas network coding has been used in V2X communication earlier \cite{ZLS}, to the best of our knowledge this is the first work that exploits index coding for V2V communication. The main contributions of this paper are as follows:
\begin{itemize}
\item A lower bound is derived for the total number of transmissions during the V2V phase, when each vehicle possesses the same number of packets. This lower bound is applicable for a cooperative data exchange problem in which each user has the same number of packets as side-information (Section \ref{sec:LB}).

\item An achievable index coding scheme is proposed for the case when there is an equal number of packets possessed in common between the adjacent vehicles. The proposed scheme meets the lower bound derived in Section \ref{sec:LB} and hence is optimal (Section \ref{sec:Enc_mtx}).

\item For the special case considered in Section \ref{sec:Enc_mtx}, we assume that the links between the vehicles are error-prone. An optimal error correcting index coding scheme is proposed for this set-up (Section \ref{sec:err_corr}).

\item  The cooperative data exchange algorithm in \cite{RSS} is applied to the case of the V2V communication and the improvement over the existing uncoded scheme in \cite{ZLS} is characterized. The proposed scheme in Section \ref{sec:Enc_mtx} is shown to perform better than the cooperative data exchange algorithm in \cite{RSS} (Section \ref{sec:simulation}).

\item Simulation results illustrating the advantage of index coding in the V2V scenario is presented (Section \ref{sec:simulation}). 
\end{itemize}

The data exchange during the V2V phase can be considered as a special case of the cooperative data exchange problem of \cite{RSS} and \cite{MPRGR}. In both, the authors study the problem of data exchange, where a group of users demands a common file, and each user has some prior information about it. In \cite{RSS}, authors provide lower and upper bounds on number of transmissions and a deterministic algorithm for the data exchange problem. However, they did not address the issue of the distribution of transmission load among the users. In \cite{MPRGR}, the authors provide the capacity region of the cooperative data exchange problem; the total transmission rate is not explicitly specified, though. A deterministic polynomial-time algorithm to find the transmission rate of individual users and a code construction based on it are developed in \cite{MPRGR}. The complexity of the algorithm increases with number of users and total number of messages. Moreover, since the code construction is based on network coding solutions, the underlying field size should be large enough to guarantee the existence of such solutions. In this work,  a closed-form solution to meet the optimum number of transmissions required for the equal overlap case is proposed. An explicit code construction to achieve the optimum rate is also developed. Our proposed code construction can operate on any field, including binary, irrespective of the number of users (vehicles) in the cluster and the total number of messages.”

\textit{Notations}: In this paper $\mathbb{F}_{q}$ denotes the finite field with $q$ elements, where $q$ is a power of a prime. The notation $[K]$ is used for the set $\{1,2,...,K\}$ for any positive integer $K$. Also $\textbf{I}_l$ denotes an $l\times l$ identity matrix, $^{\left[j \right] }\textbf{I}_l$ denotes the first j rows of $\textbf{I}_l$ and $\textbf{I}_l^{\left[j \right] }$ denotes the last j rows of $\textbf{I}_l$.  The support of a vector $\textbf{u}$ $\in \mathbb{F}^{n}_{q}$ is the set $\{i\in [n]: u_i \neq 0  \}$ and  $\textbf{e}_i$ $= (\underbrace{0,...,0}_{i-1},1,\underbrace{0,...,0}_{n-i}) \in \mathbb{F}^{n}_{q}$ denotes the unit vector having a one at the $i$th position and zeros elsewhere.

%%%%%%%%%%%%%%%%%%%%%%%%%%%%%%%%%%%%%%%%%%%%%%%%%%%%%%%%%%%%%%%%%%%%%
\section{Preliminaries and Background}

In this section, we review the collaborative message dissemination protocol in \cite{ZLS}. We also revisit some important results from index coding problem \cite{BiK} and error correcting index coding \cite{DSC}.
%%%%%%%%%%%%%%%%%%%%%%%%%%%%%%%%%%%%%%%%%%%%%%%%%%%%%%%%%%%%%%%%%%%%%%%%
\subsection{Collaborative message dissemination in VANET }
\begin{figure}
	\centering
	\includegraphics[scale=.6]{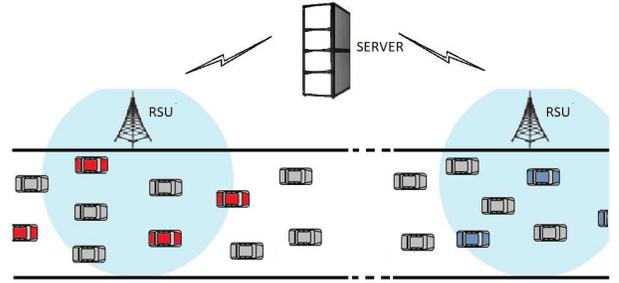}
	\caption{Scenario of collaborative data dissemination in VANET.}
	\label{fig:Scenario}
\end{figure}

The general collaborative message dissemination in VANET is motivated by the scenario where there is a multi-lane road with RSUs which are located sparsely on the roadside and there is a cluster of vehicles moving in the same direction in close communication range with each other, and this cluster of vehicles is interested in the popular content.  
This scenario is depicted in Figure \ref{fig:Scenario}, where the vehicles marked in red colour forms a cluster. The required contents are downloaded at the RSUs from the server, and the RSUs use a collaborative distribution protocol for the content dissemination. The content distribution follows three phases \cite{ZLS}. Completing the three phases is described as a round. These rounds are described as follows: 
\begin{itemize}
	\item \textit{Reporting Phase}: Vehicles report their interest to the server and the file is transferred to RSUs in the direction of the cluster. 
	\item \textit{RSU to Vehicle (R2V) phase}: In this phase, each RSU broadcasts a part of the interested file, which is not received by the vehicles in the previous rounds. Each vehicle downloads some of these packets depending on its encountering time with the RSU.
	\item \textit{Vehicle to Vehicle (V2V) phase}: In this phase, the downloaded contents are shared between vehicles after the original broadcast from the RSU. This phase is usually done when the vehicles are in the gap between two RSUs.
\end{itemize}

The number of rounds required to send the complete file from the infrastructure to the vehicles is a significant consideration of interest in this work.
The authors in \cite{ZLS}  propose three schemes to implement the protocol, namely Random-based scheme, Feedback (FB)-based scheme and Network coding (NC)-aided scheme. In the random-based scheme, vehicles do not report their received packets to the next RSU. Thus packets are randomly selected by each RSU from a set of complete packets, without knowing the packets received in each round by the cluster. 
In FB-based scheme received packets are reported to the next RSU and the packets are picked by the serving RSU only from the unreceived packets. Each RSU uses random linear network coding in the NC-aided scheme. Here, each RSU sends a linear combination of all the packets in each transmission, where the combining coefficients are selected uniformly over the finite field $\mathbb{F}_{q}$. 
%%%%%%%%%%%%%%%%%%%%%%%%%%%'
\subsection{Index Coding Problem}
Index coding problem, $\mathcal{I}$, introduced in \cite{BiK} consists of a source possessing $n$ messages $(x_1,x_2, \ldots, x_n)$, where each $x_i \in \mathbb{F}_q$ for all $i$, and $K$ receivers demanding a subset of these messages and possessing a different subset of messages as side-information. The goal is to minimize the number of coded transmissions from the source to satisfy the demands of each receiver. A receiver demanding multiple messages can be viewed as multiple receivers demanding a message each. The index of the message demanded by the $i$-th receiver is denoted by $f(i)$, where $f$ is a map from $[K]$ to $[n]$. The indices of messages possessed by the $i$-th receiver as side-information is denoted by $\mathcal{X}_i$. There are two parameters of the index coding problem which we are interested in, the min-rank $(\kappa_q(\mathcal{I}))$ and the generalized independence number $(\alpha(\mathcal{I})).$
The min-rank of an index coding problem, $\mathcal{I}$ over $\mathbb{F}_q$ is defined in \cite{DSC} as, $$
\kappa_q(\mathcal{I}) \triangleq
\text{min}\{\text{rank}_q(\{\textbf{v}_i
+\textbf{e}_{f(i)}\}_{i\in [K]}):
\textbf{v}_i \in \mathbb{F}^{n}_q, \textbf{v}_i \triangleleft \mathcal{X}_i \},$$
where $\textbf{v}_i$ $\triangleleft$ $\mathcal{X}_i$ denotes that the support of $\textbf{v}_i$ is a subset of $\mathcal{X}_i$ and $\textbf{e}_{i}$ denotes a unit vector, $(0, ..., 0 , 1, 0, ..., 0 ) \in \mathbb{F}^n_q$, having a one
at the $i^{th}$ position and zeros elsewhere. The min-rank $(\kappa_q(\mathcal{I}))$ gives the optimal length of a \textit{scalar linear index code} \cite{DSC}.

The definition of generalized independence number $(\alpha(\mathcal{I}))$ is developed as follows \cite{DSC}. The set of interfering messages is defined as $\mathcal{Y}_i=[n] \setminus \{\{f(i)\}\cup \mathcal{X}_i\}.$ A set $\mathcal{J}(\mathcal{I})$ is defined as 
\begin{equation}
\label{eq:setJ}
\mathcal{J}(\mathcal{I})= \bigcup_{i \in [K]} \{ \{f(i)\} \cup Y_i: Y_i \subseteq \mathcal{Y}_i \}. 
\end{equation}
A subset $H$ of $[n]$ is called a generalized independent set if all the non-empty subsets of $H$ lie in $\mathcal{J}(\mathcal{I})$. The maximum cardinality of $H$ is called the generalized independence number $(\alpha(\mathcal{I})).$ In general, $(\alpha(\mathcal{I}))$ serves as a lower bound for $(\kappa_q(\mathcal{I}))$ \cite{KTR}. 

Error correction for index coding with side-information was proposed in \cite{DSC}. The broadcast link between the source and the receivers is assumed to be error-prone. An error correcting index code (ECIC) which is capable of correcting at most $\delta$ errors is denoted as $\delta$-ECIC. Using an ECIC, each receiver will be able to decode its required message even if the transmissions are subjected to at most $\delta$ errors. For incorporating error correction, the source will have to make more number of transmissions. The optimal length of a $\delta$-ECIC over $\mathbb{F}_q$ is denoted by $\mathcal{N}_q [\mathcal{I}, \delta].$ There are bounds, on this optimal length established in \cite{DSC}, of which we are interested in the $\alpha$-bound and the $\kappa$-bound which are as follows:
\begin{equation}
\underbrace{N_q[\alpha(\mathcal{I}), 2\delta + 1]}_{\alpha\text{-bound}} \leq  \mathcal{N}_{q}[\mathcal{I},\delta] \leq \underbrace{N_q[\kappa_q(\mathcal{I}), 2\delta + 1]}_{\kappa\text{-bound}}. \label{eq:bds}
\end{equation} 
Here $N_q[k,d]$ denotes the optimal length of a classical error correcting code over $\mathbb{F}_q$ of dimension $k$ and minimum distance $d$. The upper bound, i.e. the $\kappa$-bound is achieved by the concatenation of an optimal index code with an optimal error correcting code which corrects $\delta$ errors. This concatenation scheme is not always optimal. For index coding problems satisfying $\alpha(\mathcal{I})=\kappa_q(\mathcal{I})$, the concatenation scheme is optimal \cite{DSC}. Error correcting index coding scheme has been used already in index coding \cite{KSR} and coded caching \cite{KTR, KTR2, KDTR}, where optimal error correcting schemes are constructed by the concatenation scheme.

%%%%%%%%%%%%%%%%%%%%%%%%%%%%%%
\section{System model and Motivating Example}
%%%%%%%%%%%
\subsection{System Model}
\label{subsec:sys_model}
In this work, a typical system model of content distribution in VANET is considered. A cluster of $K$ vehicles, moving in one direction is interested in a large file of size $L$ packets. The number of vehicles in the cluster depends on the vehicle density and the demand of interested file.  For collaborative message dissemination the vehicles should be equipped with devices based on DSRC or LTE-V standard. Both in R2V and V2V phase vehicles use 5.9GHz unlicensed spectrum for communication. If the vehicles are working on LTE-V based devices then mode-4 operation can be selected. All the vehicles in the cluster should rely on either 802.11p based devices or on LTE-V based devices. Also, the vehicles in a cluster are moving with relatively close velocities. 

Vehicles download the file packets from RSU to reduce the cost associated with network usage. When these vehicles pass by the RSU, the RSU delivers a part of the file and is known as RSU to vehicle(R2V) phase. Let $n$ be the number of packets delivered by RSU in the RSU to vehicle (R2V) phase. We assume that each vehicle is able to receive $l$ number of consecutive packets delivered by RSU with $l \leq n$. Here, $l$ is known as the download capability of the vehicle and depends on the velocity of vehicles. If all the vehicles move with the same velocity, their download capability will be the same. Any two adjacent vehicles may receive some common packets as well. When the cluster leaves the RSU, vehicles communicate among themselves so that all the vehicles get all the packets delivered by the RSU. This is known as the vehicle to vehicle (V2V) phase. Several R2V and V2V phases are required to deliver the whole file to all the vehicles.

%%%%%%%%%%%%%%%%
\subsection{Index coding problem in the V2V phase}
\label{sec:IC_V2V}
Consider a set of $n$ packets, $X= \lbrace x_1,x_2,...x_n \rbrace $    transmitted by RSU to a cluster of  vehicles  $V_1,V_2,\ldots, V_K$ in the R2V phase. The vehicles in the cluster communicate among themselves in the V2V phase so that the complete set ${X}$ is received by all vehicles. This can be viewed as a cooperative data exchange problem \cite{RSS}, or as the delivery phase of a D2D coded caching scheme \cite{JCM} or as $K$ simultaneous index coding problems. We take the index coding approach in order to reduce the number of transmissions in the V2V phase.  We denote the index coding problem in a D2D set-up like this as $\mathcal{D}$ and the minimum number of transmissions required for $\mathcal{D}$ as $\kappa(\mathcal{D})$. Elements of set, $\mathcal{K}_m \subseteq {X}$, $m \in \left[ K \right] $ are packets received by the vehicles in the R2V phase such that $\cup_{{m\in \left[ K \right]}} \mathcal{K}_m= {X}$, known as side-information. The download capability of all the vehicles in the R2V phase are assumed to be same, i.e. $|\mathcal{K}_m|=l, \forall m$. Each vehicle, $V_m$ in the cluster is described by a tuple $(\mathcal{K}_m,\mathcal{W}_m)$, where the set $\mathcal{W}_m = {X} \backslash \mathcal{K}_m,$ is known as the want set of $V_m$ with $|\mathcal{W}_m|=n-l$.  The set $\mathcal{C}_m$ has the coded messages transmitted by the vehicle $V_m$ in the V2V phase. The total length of index code is given as $\sum_{m=1}^{K}{|\mathcal{C}_m|}$. 
\subsection{Motivating Example: Illustration of Index coding problem in V2V phase}
\label{sec:motivating_example}
In this section, an example is given illustrating the use of index coding (IC) in the V2V phase to reduce the number of transmissions.  
\begin{example}
\label{ex:k4l4}
Let $n=8$, $K=4$ and $l=4$. The vehicles have their received sets of messages and the corresponding want sets of messages as given in Table \ref{Illus_expl}. The index coded packets transmitted by each vehicle is given as $\mathcal{C}_1 =\lbrace x_1+x_3,x_2+x_4 \rbrace$,$\mathcal{C}_2 =\lbrace x_3+x_5\rbrace$, $\mathcal{C}_3 =\lbrace x_4+x_6 \rbrace$, and $\mathcal{C}_4 =\lbrace x_7+x_5,x_8 +x_6\rbrace.$ This is an example of general index coding problem in V2V phase when vehicles are moving at constant velocity.  For $n=8$, at least $8$ uncoded transmissions should be necessary in order for all the vehicles to receive all the packets.  The number of index coded messages in this example is, $\sum_{m=1}^{4}{|\mathcal{C}_m|}=6$. Therefore, two transmissions can be saved in each V2V in this particular example. So this example establishes that index coding is possible in the V2V phase and the number of transmissions can be saved using it. 
\end{example}

\begin{table}[]
		\centering
		\begin{tabular}{|c|c|c|}
			\hline
			vehicle & {Known set}& {Want set} \\ \hline
			$V_1 $     &$\{ x_1, x_2, x_3, x_4\} $  & $\{x_5, x_6,x_7,x_8\}$   \\ \hline
			$	V_2 $     & $\{x_3, x_4, x_5, x_6\} $ &$ \{x_1, x_2, x_7, x_8\} $  \\ \hline
			$	V_3 $     &$ \{x_4, x_5, x_6, x_7\} $       & $\{x_1, x_2, x_3, x_8\}$    \\ \hline
			$	V_4  $    &$ \{x_5,x_6,x_7,x_8\}$        & $\{ x_1, x_2, x_3, x_4\} $     \\ \hline
		\end{tabular}
		\caption{Example of IC problem  with $K=4$ and $l=4$.}
		\label{Illus_expl}
\end{table}
%%%%%%%%%%%%%%%%%%%%%%%%%
\section{Lower bound on IC transmissions in V2V phase}
\label{sec:LB}
In this section, we give a lower bound for the system model we presented in Section \ref{subsec:sys_model}.  Let $\mathcal{N}_m$ be a set of \emph{innovative packets} that $V_m$ receives. An innovative packet to a vehicle is a packet that the vehicle receives and no other vehicles receive. In Example \ref{ex:k4l4}, the innovative packets possessed by the vehicle $V_1$ are $x_1$ and $x_2$, hence $\mathcal{N}_1=\{x_1, x_2\}.$ Also, for this example, $\mathcal{N}_2$ and $\mathcal{N}_3$ are null sets and $\mathcal{N}_4= \{ x_8\}.$  
\begin{table*} 
	\centering
	\begin{tabular}{|c|c|c|}
		\hline
		Vehicle & {Known set}&{Want set} \\ \hline
		$V_1$ & $\{x_1, x_2, x_3, x_4, x_5\}$ & $\{x_6, x_7, x_8,x_9, x_{10}, x_{11}, x_{12}, x_{13}, x_{14}\}$ \\ \hline 
		$V_2$ & $\{x_4, x_5, x_6, x_7, x_8\}$ & $\{x_1, x_2, x_3,x_9, x_{10}, x_{11}, x_{12}, x_{13}, x_{14}\}$ \\ \hline 
		$V_3$ & $\{x_7, x_8, x_9, x_{10}, x_{11}\}$ & $\{x_1, x_2, x_3,x_4, x_{5}, x_{6}, x_{12}, x_{13}, x_{14}\}$ \\ \hline 
		$V_4$ & $\{x_{10}, x_{11}, x_{12}, x_{13}, x_{14}\}$ & $\{x_1, x_2, x_3, x_4, x_{5}, x_{6}, x_{7}, x_{8}, x_{9}\}$ \\ \hline 
	\end{tabular}
	\caption{Example problem  with $K=4$, $l=5$ and $i=2$}
	\label{expl2}
\end{table*}
	 \begin{theorem}
	 \label{thm:LB}
	 A lower bound on the number of transmissions required in the V2V scenario where all the vehicles have equal number of packets with them is  $n-l+\max\limits_{{m\in \left[ K \right]}} |\mathcal{N}_m|$
	 \end{theorem}
\begin{IEEEproof}
Any vehicle, $V_m$, should receive at least $n-l$ coded packets from the other vehicles in order to get all the packets of its want set.  A vehicle is not benefited from its transmissions. Also, no other vehicle has innovative packets of $V_m$. In view of the above facts,  $V_m$ should do at least $|\mathcal{N}_m|$  transmissions, which is the number of innovative packets with it, so that all the other vehicles can decode these innovative packets.     Considering the worst case, we consider the vehicle with the maximum number of innovative packets. Hence a lower bound for the optimal number of transmissions  is given by $n-l+\max\limits_{{m\in \left[ K \right]}} |\mathcal{N}_m|$. 
	 \end{IEEEproof}
%%%%%%%
\subsection{Equal overlap case}
	We consider a special case in which the vehicles are equally spaced and move at the same speed.  In this case, the number of common packets received by the adjacent vehicles will be the same under the error-free channel assumption. We call this case as the equal overlap case. Let $\mathcal{Z}_j,j\in \left[ K-1\right]$ be a set containing the common elements in $\mathcal{K}_j$ and $\mathcal{K}_{j+1}$. In this case, $|\mathcal{Z}_j|$ will be same for all vehicles and is denoted by $i,0\le i \le l$. i.e. $|\mathcal{K}_j \cap \mathcal{K}_{j+1}|=i,\quad  \forall j\in \left[ K-1\right] $. This situation is shown as an example in Table \ref{expl2},  with $K=4$,  $l=5$ and $i=2$. 

\begin{corollary}
\label{cor}
For the equal overlap case, $\kappa(\mathcal{D}) \geq n-i.$
\end{corollary}
\begin{IEEEproof}
In this case, since the last vehicle receives the last $l$ packets, we have $n = K(l-i)+i$ is the total number of packets delivered by RSU. The number of innovative packets received in this scenario is $l-2i$ by all vehicles except for the first and the last vehicle. The number of innovative packets that the first and the last vehicle receives is $l-i$. So $n-l+\max\left(l-i, l-2i \right) = n-i$ is the lower bound on the number of transmissions required in this scenario. 
\end{IEEEproof}
%%%%%%%%%%%%%%%%%%%%%%%%%%%%%%%
\section{Optimal Scheme for the Equal Overlap Case}
\label{sec:Enc_mtx}
For the equal overlap case, we propose an achievable scheme for transmission, which in turn meets the lower bound in Corollary \ref{cor}. Let $\textbf{I}_l$ denotes an $l\times l$ identity matrix, $^{\left[j \right] }\textbf{I}_l$ denotes the first j rows of $\textbf{I}_l$ and $\textbf{I}_l^{\left[j \right] }$ denotes the last j rows of $\textbf{I}_l$. Then the encoding matrix used by the vehicle $V_m$ is given as 
\begin{equation}
\textbf{L}=\textbf{I}_l^{\left[l-i \right] }+^{\left[l-i \right] }\textbf{I}_l
\label{eq1}
\end{equation}

Let $\textbf{x}_m=\left(x_{m,1},x_{m,2},...,x_{m,l} \right)^T $ be an $l \times 1$ vector representing the messages received by $V_m$. The index coded packets broadcast by $V_m$ is given by $\textbf{c}_m =\textbf{L}\textbf{x}_m$, where $\textbf{c}_m$ is an $l-i\times 1$ vector of the form $\left( c_{m,1},c_{m,2},..,c_{m,l-i} \right)^T $.  
This encoding matrix is the same for all the vehicles. Hence each vehicle uses $l-i$ transmissions. The total number of transmissions is thus $K(l-i)$. From the constraint $n=K(l-i)+i$, we have $K(l-i)=n-i.$ Hence this meets the lower bound in the Corollary \ref{cor} and is optimal. This is illustrated using an example given below.

\begin{example}
 Consider the case shown in the example in Table \ref{expl2} with $K=4$,  $l=5$ and $i=2$. So here $l-i=3$ transmissions are made by each vehicle. The matrices  $^{\left[l-i \right] }\textbf{I}_l$ and $ \textbf{I}_l^{\left[l-i \right] }$ are given as: 
 $$  ^{\left[l-i \right] }\textbf{I}_l=\left[ {\begin{array}{lllll}
        1 & 0 & 0 & 0 & 0 \\
        0 & 1 & 0 & 0 & 0 \\
    
        0& 0 & 1 & 0 & 0  \\

        \end{array} } \right], $$
$$     \textbf{I}_l^{\left[l-i \right] } =\left[ {\begin{array}{lllll}
            0 & 0 & 1 & 0 & 0 \\
            0 & 0 & 0 & 1 & 0 \\
        
            0& 0 & 0 & 0 & 1  \\
    
            \end{array} } \right].   
    $$ 
  As per \eqref{eq1} we have, 
   \begin{equation}   
   \textbf{L}=\left[ {\begin{array}{lllll}
               1 & 0 & 1 & 0 & 0 \\
               0 & 1 & 0 & 1 & 0 \\
           
               0& 0 & 1 & 0 & 1  \\
       
               \end{array} } \right].
           \label{eq2}
    \end{equation} 
               For $m=2$ and $\textbf{x}_2 = (x_{4},x_{5},x_{6},x_{7},x_{8})^T $, $\textbf{c}_2$ is given as
               \begin{equation}   
               \textbf{c}_2=\left[ {\begin{array}{lllll}
                              1 & 0 & 1 & 0 & 0 \\
                              0 & 1 & 0 & 1 & 0 \\
                          
                              0& 0 & 1 & 0 & 1  \\
                      
                              \end{array} } \right]\left[ {\begin{array}{l}
                                                            x_{4}  \\
                                                            x_{5}  \\
                                                            x_{6}  \\
                                                            x_{7}  \\
                                                            x_{8}  \\

                                                            \end{array} } \right] 
                           =\left[ {\begin{array}{l}
                           x_{4}+ x_{6} \\
                           x_{5}+ x_{7}  \\
                           x_{6}+ x_{8}  \\
                           \end{array} } \right]. 
                          \label{eq3}
                     \end{equation}
\end{example}
%%%%%%%%%%%%%
\subsection{Decodability}
In this subsection, we show that using the proposed code, each vehicle decodes all the messages. Towards this, we prove initially that from the transmissions of vehicle $V_m$, the vehicles $V_{m-1}$ and $V_{m+1}$ are able to decode all the messages possessed by $V_m$. Using this, we prove that all the vehicles are able to decode all the messages which they do not possess. Also, we give an example to illustrate the decodability.  

\begin{lemma}
	 \label{lem:rtdec}
$V_{m-1}$ can decode $\textbf{x}_m$ from the transmissions of $V_m.$
\end{lemma}
\begin{IEEEproof}
%$\textbf{x}_m$ decodability of $V_{m-1}$ implies $V_m$'s coded transmissions, $C_m^{l-i}$ is sufficient to decode the $\textbf{x}_m$ vector by $V_{m-1}$.  
Define an $i\times 1$ vector $\textbf{z}_{m-1,m} $ as the side information vector of $V_{m-1}$ with the elements as the members of the set $\mathcal{K}_{m}\cap \mathcal{K}_{m-1}$. By definition, $\textbf{z}_{m-1,m} $ contains the first $i$ elements of $\textbf{x}_m$ or the last $i$ elements of $\textbf{x}_{m-1}$. So $\textbf{z}_{m-1,m}=\left(x_{m,1},x_{m,2},...,x_{m,i} \right)^T $.
Let $\textbf{P}_R$ be a decoding matrix for $V_{m-1}$ defined as
\begin{equation}  
 \textbf{P}_R.\textbf{x}_m =\left[ {\begin{array}{l}
	\textbf{z}_{m-1,m}  \\
	\textbf{c}_m \\
	
	\end{array} } \right].
\label{eq5}
\end{equation}
Then,
\begin{equation}   
\textbf{P}_R =\left[ {\begin{array}{c}
            ^{\left[i \right] }\textbf{I}_l  \\
            \textbf{L} \\
    
            \end{array} } \right]_{l\times l}.
        \label{eq4}
   \end{equation}
The messages in $\textbf{x}_m$ can be decoded by $V_{m-1}$, if $\textbf{P}_R$ is a full rank matrix. From \eqref{eq1} and \eqref{eq4} $\textbf{P}_R$ can be written as
      \begin{equation} 
      \textbf{P}_R =\textbf{I}_l+\left[ {\begin{array}{c}
                  \textbf{0}_{i \times l}  \\
                  ^{\left[l-i \right]}\textbf{I}_l \\
          
                  \end{array} } \right],
              \label{eq6} 
       \end{equation}
where $\textbf{0}_{i \times l}$ is an $i \times l$ all zero matrix. Since $^{\left[l-i \right]}\textbf{I}_l$ is the first $l-i$ rows of $\textbf{I}_l$, the reduced row echelon form (rref) of $\textbf{P}_R$ will be  $\textbf{I}_l$. Hence $\textbf{P}_R$ is a full rank matrix and $V_{m-1}$  can decode the messages in $\textbf{x}_m$.
 \end{IEEEproof}   

The decodability explained in Lemma \ref{lem:rtdec}  is illustrated in the example given below.
%%%%%%%%%%
  \begin{example}
Consider the example given in Table \ref{expl2}. We have,  
  
  $$ \mathbf{z}_{1,2}=\left[ {\begin{array}{l}
  	x_{4} \\
  	x_{5}  \\
  	\end{array} } \right], 
  	\textbf{P}_R=\left[ {\begin{array}{c}
   	^{\left[2 \right] }\textbf{I}_5  \\
   	\textbf{L} \\
   	
   	\end{array} } \right]=\left[ {\begin{array}{lllll}
   	1 & 0 & 0 & 0 & 0 \\
   	0 & 1 & 0 & 0 & 0 \\
   	1& 0 & 1 & 0 &0  \\	
   	0&1 & 0 & 1 & 0  \\
   	0& 0 & 1 & 0 & 1  \\
   	\end{array} } \right].
  	$$
%  \begin{equation}  $$\begin{center} $  \textbf{c}_2^{3}=\left[ {\begin{array}{l}
%  	x_{4}+ x_{6} \\
%  	x_{5}+ x_{7}  \\
%  	x_{6}+ x_{8}  \\
%  	\end{array} } \right] $
%  \end{center}
%  \label{eq32}
%  $$\end{equation}    
%     
   Hence, we have
     \begin{equation} \nonumber $$\begin{center} $  \textbf{P}_R\textbf{x}_m=\left[ {\begin{array}{lllll}
   	1 & 0 & 0 & 0 & 0 \\
   	0 & 1 & 0 & 0 & 0 \\
   	1& 0 & 1 & 0 & 0  \\	
   	0&1 & 0 & 1 & 0 \\
   	0& 0 & 1 & 0 & 1  \\
   	
   	\end{array} } \right]\left[ {\begin{array}{l}
   	x_{4}  \\
   	x_{5}  \\
   	x_{6}  \\
   	x_{7}  \\
   	x_{8}  \\

   	\end{array} } \right] 
   =\left[ {\begin{array}{l}
   	x_{4} \\
   	x_{5}  \\
   	x_{4}+ x_{6} \\
   	x_{5}+ x_{7}  \\
   	x_{6}+ x_{8}  \\
   	\end{array} } \right]. $
   \end{center}
   \label{eq34}
   $$\end{equation}
      Applying row transformations on  $\textbf{P}_R$ as $R_3\rightarrow R_3-R_1$, $R_4\rightarrow R_4-R_2$ and $R_5\rightarrow R_5-R_3$, this gives the row-reduced echelon form, denoted by $rref(\textbf{P}_R)$ as:
       
       \begin{equation} \nonumber $$\begin{center} $  rref(\textbf{P}_R)=\left[ {\begin{array}{lllll}
       	1 & 0 & 0 & 0 & 0 \\
       	0 & 1 & 0 & 0 & 0 \\
       	0& 0 & 1 & 0 & 0  \\	
       	0&0 & 0 & 1 & 0 \\
       	0& 0 &0 & 0 & 1  \\
       	
       	\end{array} } \right].$
       \end{center}
       \label{eq35}
       $$\end{equation}
This means that $\textbf{P}_R$ is full rank. 
 \end{example} 
  
\begin{lemma}
 	 \label{lem:ltdec}
 $V_{m+1}$ can decode $\textbf{x}_m$ from the transmissions of $V_m.$
 \end{lemma}
 \begin{IEEEproof}
% $\textbf{x}_m$ decodability of $V_{m+1}$ implies that $V_{m+1}$ can decode the vector $\textbf{x}_m$ with $C_m^{l-i} $ . 
The vector $\textbf{z}_{m,m+1} $ has the elements of the set $\mathcal{K}_{m}\cap \mathcal{K}_{m+1}$ and is the side information vector of $V_{m+1}$. So $\textbf{z}_{m,m+1}=\left(x_{m,l-i+1},...x_{m,l-1},x_{m,l} \right)^T $.
Let $\textbf{P}_L$ be a decoding matrix of $V_{m+1}$  defined as
 \begin{equation}   
 \textbf{P}_L =\left[ {\begin{array}{c}
             \textbf{L} \\
     		\textbf{I}_l^{\left[i \right] }  \\
             \end{array} } \right]_{l\times l}.
         \label{eq7}
   \end{equation}
Then,
 \begin{equation}   
 \textbf{P}_L.\textbf{x}_m =\left[ {\begin{array}{l}
             
             \textbf{c}_m \\
       		\textbf{z}_{m,m+1}  \\
             \end{array} } \right].
         \label{eq8}
   \end{equation}
The messages in $\textbf{x}_m$ can be decoded by $V_{m+1}$, if $\textbf{P}_L$ is a full rank matrix. From \eqref{eq1} and \eqref{eq7} $\textbf{P}_L$ can be written as
       \begin{equation}   
       \textbf{P}_L =\textbf{I}_l+\left[ {\begin{array}{c}
                   \textbf{I}_l^{\left[l-i \right]} \\
           			\textbf{0}_{i \times l}  \\
                   \end{array} } \right].
               \label{eq9} 
         \end{equation}
 Since $\textbf{I}_l^{\left[l-i \right]}$ is the last $l-i$ rows of $\textbf{I}_l$, the reduced row echelon form of $\textbf{P}_L$ will be  $\textbf{I}_l$. Hence $\textbf{P}_L$ is a full rank matrix and $V_{m+1}$  can decode the messages in $\textbf{x}_m$.     
 \end{IEEEproof}
The decodability in Lemma \ref{lem:ltdec} is illustrated in the example below.
\begin{example}
Decodability in Lemma \ref{lem:ltdec} is illustrated for the example in Table  \ref{expl2}. We have, 
$$ 
\mathbf{z}_{2,3}=\left[ {\begin{array}{l}
	x_{7} \\
	x_{8}  \\
	\end{array} } \right],
	\textbf{P}_L=\left[ {\begin{array}{c}
	
	\textbf{L} \\
	\textbf{I}_5^{\left[2 \right] }  \\
	\end{array} } \right]=\left[ {\begin{array}{lllll}
	1 & 0 & 1 & 0 & 0 \\
	0 & 1 & 0 &1 & 0 \\
	0& 0 & 1 & 0 &1  \\	
	0&0 & 0 & 1 & 0  \\
	0& 0 & 0 & 0 & 1  \\
	\end{array} } \right],
$$   
\begin{equation} \nonumber $$\begin{center} $  \textbf{P}_L\textbf{x}_m=\left[ {\begin{array}{lllll}
		1 & 0 & 1 & 0 & 0 \\
	0 & 1 & 0 &1 & 0 \\
	0& 0 & 1 & 0 &1  \\	
	0&0 & 0 & 1 & 0  \\
	0& 0 & 0 & 0 & 1  \\
	
	\end{array} } \right]\left[ {\begin{array}{l}
	x_{4}  \\
	x_{5}  \\
	x_{6}  \\
	x_{7}  \\
	x_{8}  \\

	\end{array} } \right] 
=\left[ {\begin{array}{l}
	
	x_{4}+ x_{6} \\
	x_{5}+ x_{7}  \\
	x_{6}+ x_{8}  \\
	x_{7} \\
	x_{8}  \\
	
	\end{array} } \right].$
\end{center}
\label{eq39}
$$\end{equation}
Applying row transformations on  $\textbf{P}_L$ as $R_1\rightarrow R_1-R_3$, $R_2\rightarrow R_2-R_4$ and $R_3\rightarrow R_3-R_5$, we get the row-reduced echelon form, denoted by $rref(\textbf{P}_L)$ as:
\begin{equation} \nonumber  $$\begin{center} $  rref(\textbf{P}_L)=\left[ {\begin{array}{lllll}
	1 & 0 & 0 & 0 & 0 \\
	0 & 1 & 0 & 0 & 0 \\
	0 & 0 & 1 & 0 & 0  \\	
	0 & 0 & 0 & 1 & 0 \\
	0& 0 &0 & 0 & 1  \\
	\end{array} } \right].$
\end{center}
\label{eq310}
$$\end{equation}
  \end{example} 
%\subsection{Complete decodability}
%Complete decodability ensures that all the vehicles can get their desired packets. Encoding matrix $L$ is the same for all the vehicles,  therefore neighboring decodability is guaranteed for all the vehicles. 
\begin{theorem}
\label{thm:dec}
If each vehicle transmits $\textbf{c}_m =\textbf{L}\textbf{x}_m$, where $\textbf{L}$ is as given in \eqref{eq1}, all the vehicles are able to decode all the messages in the equal overlap case.
\end{theorem}
\begin{IEEEproof}
In Lemma \ref{lem:rtdec} and Lemma \ref{lem:ltdec} we have shown that $V_m$ can get the messages in $\mathcal{K}_{m-1}$ and $\mathcal{K}_{m+1}$ with $\mathcal{K}_{m}$ as side information. After decoding, using $\mathcal{K}_{m-1}$ as side information $V_m$ will decode $\mathcal{K}_{m-2}$ and with $\mathcal{K}_{m+1}$ as side information $V_m$ will decode $\mathcal{K}_{m+2}$. This process goes on until all the packets can be decoded by $V_m$. It refers to any value of m, thus ensuring the total decodability. 
 \end{IEEEproof}
 %%%%%%%%%%%%%%%%%%
\section{Error Correction for the Equal Overlap Case}
\label{sec:err_corr}
In this section, we consider the case that the transmissions are prone to errors. To construct an optimal error-correcting index code, we find the generalized independence number for all the index coding problems as viewed by all the vehicles as they transmit. Let the index coding problem seen when the vehicle $V_m$ transmits be $\mathcal{I}^m$.  Besides, we consider that for the index coding problem $\mathcal{I}^m$, there are only at most two receivers, which are the vehicles $V_{m-1}$ and $V_{m+1}$. Note that for the index coding problem $\mathcal{I}^1$, there will be only one receiver, i.e., the vehicle $V_2$. Also, for the index coding problem $\mathcal{I}^K$, the only receiver is $V_{K-1}.$ We assume that when the vehicle $V_m$ transmits, the messages in the index coding problem   $\mathcal{I}^m$ solely consists of the messages which are possessed by the vehicle $V_m$.
The demanded messages and the side-information of the receiver vehicles are modified as $\mathcal{W}_i^m= \mathcal{W}_i \cap \mathcal{K}_m$ and  $\mathcal{K}_i^m= \mathcal{K}_i \cap \mathcal{K}_m$. This is illustrated using an example as shown below.
\begin{example}
Consider the V2V setting in Example \ref{ex:k4l4}. The index coding problem corresponding to the transmissions of $V_1$, $\mathcal{I}^1$, has the message set consisting of the messages in $\mathcal{K}_1 = \{x_1, x_2, x_3, x_4 \}$. The only receiver $V_2$, will have the demand set and the side-information set given as: 
$$\mathcal{K}_2^1=\{x_3, x_4 \}, \mathcal{W}_2^1=\{x_1, x_2\}, $$
For the index coding problem $\mathcal{I}^2$, the message set is $\mathcal{K}_2 = \{x_3, x_4, x_5, x_6 \}$. The demand and side-information sets for the corresponding receivers $V_1$ and $V_3$ are as follows.
$$\mathcal{K}_1^2=\{x_3, x_4\}, \mathcal{W}_1^2=\{x_5, x_6\}, $$
  $$\mathcal{K}_3^2=\{x_4, x_5\}, \mathcal{W}_3^2=\{x_3\}. $$
Further, each receiver demanding more than one message can be viewed as multiple receivers demanding one message each.
\end{example}

From Lemma \ref{lem:rtdec} and Lemma \ref{lem:ltdec}, it is ensured that the receivers in $\mathcal{I}^{m}$ can decode their demanded messages. Hence, a vehicle acts as a receiver in at most index coding problems.  For instance, vehicle $V_m$ acts as a receiver in index coding problems $\mathcal{I}^{m-1}$ and $\mathcal{I}^{m+1}$. This receiver thus directly decodes all the messages which $V_{m-1}$ and $V_{m+1}$ have. Even though it is not a receiver in any other index coding problems, because of the broadcast nature, it listens to all the other transmissions also and further decodes all the other packets it wants, which follows from the proof of Theorem \ref{thm:dec}.

 We denote the generalized independence number and the min-rank of the index coding problem $\mathcal{I}^m$ as $\alpha(m)$ and $\kappa(m)$ respectively.
\begin{lemma}
\label{lem_alpha}
The generalized independence number of the index  coding problem $\mathcal{I}^{m}$ satisfies:
$\alpha(m) \geq (l-i).$
\end{lemma}
\begin{IEEEproof}
We construct a generalized independent set $B(m)$, whose cardinality will be a lower bound on $\alpha(m)$. The set which we construct is:
$$ B(m) = \{x_{(m-1)l+1},x_{(m-1)l+2}, \ldots, x_{(m-1)l+(l-i)}\}.$$
The claim is that the indices of elements of $B(m)$ form a generalized independent set. All the messages corresponding to the messages in $B(m)$ are demanded, hence all the singleton subsets of $B(1)$ are present in the set $\mathcal{J}(\mathcal{I}^m)$ as defined in \eqref{eq:setJ}. Since all the messages of $B(m)$ are  not possessed by the vehicle $V_{m+1}$, all the non-empty subsets of $B(m)$ are present in the set $\mathcal{J}(\mathcal{I}^m).$ Thus $B(m)$ is a generalized independent set and $\alpha(m) \geq (l-i).$  
\end{IEEEproof}
\begin{theorem}
For the equal overlap case, $\alpha(m)= \kappa(m),$ for all $m \in [K].$ 
\end{theorem}
\begin{IEEEproof}
From the optimal scheme explained in Section \ref{sec:Enc_mtx}, we have seen that each vehicle makes $(l-i)$ transmissions. Hence, $\kappa(m) \leq (l-i).$ From \cite{DSC} and \cite{KTR}, $\alpha(m) \leq \kappa(m).$ Also from Lemma \ref{lem_alpha}, we have $\alpha(m)= \kappa(m)$ for all $m \in [K].$
\end{IEEEproof}

From the definition of $\kappa_q(\mathcal{I})$ for an index coding problem $\mathcal{I}$, $\kappa_q(\mathcal{I})$ is a non-increasing function of the field size $q$ \cite{DSC}. 
The transmissions corresponding to the proposed scheme in Section \ref{sec:Enc_mtx} is over the binary field. Hence the value of $\kappa(m)$ is found over the binary field. Besides, since $\alpha(m)= \kappa(m),$ the value of $\kappa(m)$ can get no better even if a larger field is considered. Since for all the vehicles, we have $\alpha(m)= \kappa(m),$ the $\alpha$ and $\kappa$ bounds in \eqref{eq:bds} meet with equality. Hence the optimal error-correcting scheme employed at each vehicle is by concatenating the optimal scheme in Section \ref{sec:Enc_mtx} with an optimal error-correcting code that corrects the required number of errors. The total number of transmissions which will enable each vehicle to incorporate error-correcting capability of $\delta$ is thus, $K N_q[\alpha(m), 2\delta+1].$ From the transmissions of $V_m$, vehicles $V_{m-1}$ and $V_{m+1}$ can decode the demanded messages without errors. Following the arguments from the proof of Theorem \ref{thm:dec}, all vehicles can decode all the required packets without error. 
This is illustrated by the following example:
\begin{example}
Consider the equal overlap scenario considered in Table \ref{expl2} with $K=4$, $l=5,$ $i=2,$ and the field size $q=2.$ For each vehicle $V_m$ we have $\alpha(m)=3.$ Moreover, by the optimal scheme presented in Section \ref{sec:Enc_mtx}, we have $\kappa(m)=3.$ Hence, we have $\alpha(m)= \kappa(m)=3$ for all the vehicles. Consider the case when $\delta=1$ error needs to be corrected by each vehicle during its transmission. Since the $\alpha$ and $\kappa$ bounds meet, the optimal error-correcting scheme here is to concatenate the optimal code in Section \ref{sec:Enc_mtx} with an optimal single error-correcting code. For binary field, we have $N_2[3,3]=6$ from \cite{Gra}. Hence, the concatenation can be done with a $[6,3,3]_2$ code. 
A generator matrix corresponding to $[6,3,3]$-code is
$$
\textbf{G}=
\begin{bmatrix}
1 & 0 & 0 & 1 & 1 & 0 \\
0 & 1 & 0 & 1 & 0 & 1 \\
0 & 0 & 1 & 0 & 1 & 1
\end{bmatrix}.
$$ 
Hence each vehicle makes $6$ transmissions. Total number of transmissions here will be $6 \times 4 =24.$
\end{example}
%%%%%%%%%%%%%
\section{Simulation results}
\label{sec:simulation}
In perfect V2V sharing, complete data transfer is assumed, i.e. all the $K$ vehicles can exchange $n$ packets among them. It is clear from the previous discussions that index coding at V2V phase can reduce the number of transmissions for perfect V2V. In this section, we present some simulation results by assuming imperfect V2V.  In imperfect V2V, the number of data exchange is less than the minimum required number. Imperfect V2V may happen if the spacing between the RSUs is not enough for a perfect V2V exchange, i.e. when the vehicles enter the coverage of the next RSU, they stop V2V sharing and start downloading from the next RSU. Any of the vehicles leaving the cluster before perfect V2V transfer, also can result in imperfect V2V. 

 The parameter $l$ refers to the download capability of a vehicle when it passes by the RSU. This parameter depends on the data rate of RSU transmissions, speed of the vehicle and the coverage area of RSU. If the transmission rate is higher, speed is lower, and RSU covers a larger area, then $l$ could be higher.  For example, if the speed of the vehicle is $20~m/s$ and length of road covered by RSU is $400~m$, then for a transmission rate of $2~Mbps$ and a packet size of $2.5~MB$, the download capability is $l=4$. The number of vehicles in the cluster ($K$) depends on the vehicle density and the popularity of the file. The total number of packets that need to be exchanged in V2V phase ($n$) mainly depends on the RSU coverage, transmission rate and the size of the cluster. Size of the cluster refers to the distance between the first and the last vehicle.
We use matlab set up for simulations and for graphs each simulation is repeated 100 times. Before illustrating the simulation results on the index coding techniques, we present the results when the cooperative data exchange algorithm in \cite{RSS} is used in this V2V scenario. 
%%%%%%%%%%%%%%%
\subsection{Cooperative data exchange \cite{RSS}}
\label{sec:sadeghi_alg}
The problem of cooperative data exchange was introduced in \cite{RSS}, where $K$ users demand the whole message set and each user possesses a subset of the message as side information set. Upper and lower bounds on the minimum number of transmissions are established in \cite{RSS}. From \cite{RSS}, we have, $\kappa(\mathcal{D}) \geq n-n_{min}$, where $n_{\text{min}}$  denote the minimum cardinality of the side-information set. For the case when all the users have the same number of messages as side-information, i.e., $|\mathcal{K}_i| = n_{min}$ $ \forall i$,  $\kappa(\mathcal{D}) \geq n-n_{min}+1.$ An upper bound on minimum required transmissions is given by $\kappa(\mathcal{D}) \leq \min\limits_{1\leq i\leq K} \{{|\mathcal{K}_i|+\max\limits_{1\leq j\leq K}|\overline{\mathcal{K}}_j \cap \mathcal{K}_i|}\}$. Algorithmic solution for coded message dissemination in this setting is given in \cite{RSS}, which uses a near-optimal number of broadcasts. We reproduce this algorithm here as Algorithm 1, where  the vector $\mathbf{\Gamma}_c$ denotes the linear coefficients associated with a coded packet $\textbf{c}$, such that $\textbf{c}=\mathbf{\Gamma}_c.(x_1 x_2 \ldots x_n)^T$. We implement this algorithm in the V2V set-up and show that this gives considerable improvement over the scheme in \cite{ZLS}.

\begin{algorithm}
	\SetKwData{Left}{left}
	\SetKwData{Up}{up}
	\SetKwFunction{FindCompress}{FindCompress}
	\SetKwInOut{Input}{input}
	\SetKwInOut{Output}{output}
	\Indm\Indmm
	\Indp\Indpp
	\DontPrintSemicolon
	\BlankLine
	 $U=[K]$.
	\BlankLine
	\For{$i$=1 to $K$}    
	{ 
		$Y_i=\langle\{ \mathbf{\Gamma}_x,x\in \mathcal{K}_i\}\rangle$, span of the set $\{ \mathbf{\Gamma}_x,x\in \mathcal{K}_i\}$
	}
	\BlankLine
	\While{there is a user i with $dim(Y_i) < n$}
	{
		\While{$\exists i, j\in U, i \ne j$ such that $Y_i=Y_j$ }
		{
			$U=U \setminus i$\;
		}
		
		\BlankLine
		Find a user $i$ with a subspace $Y_i$ of maximum dimension (If there are multiple
		such clients choose one of them arbitrarily)
		\BlankLine
		Select a vector $\textbf{b} \in Y_i$ such that $\textbf{b} \notin Y_j$
		for each $j \ne i$
		\BlankLine
		User $i$ broadcast the packet $\textbf{c}=\textbf{b}.(x_1 x_2 \ldots x_n)^T$
		\BlankLine
		\For{$i$=1 to $K$}    
		{ 
			$Y_i=Y_i + \langle\{ \textbf{b} \}\rangle$
		}
		\BlankLine
	}
	\Indm\Indmm
	\Indp\Indpp
	\caption{Algorithm for Information Exchange \cite{RSS}}
\end{algorithm}

In collaborative message dissemination scenario under consideration, the $|\mathcal{K}_m|$ is considered the same for different vehicles. Algorithm for cooperative data exchange is simulated for this scenario, and the results are as shown in Table \ref{AlgSimLB}.
A gap between the actual number of transmissions required by the algorithm and the lower bound provided is identified. 
\begin{table}[t]
	\centering
	\begin{tabular}{|l|l|l|l|l|l|}
		\hline
		Total number of packets, $n$ & 10 & 20 & 30 & 40 & 50 \\ \hline
		Download capability, $l$ & 6  & 8  & 10 & 12 & 14 \\ \hline
		Lower bound \cite{RSS}         & 5  & 13 & 21 & 29 & 37 \\ \hline
		Number of transmissions & 5  & 15 & 25 & 36 & 46 \\ \hline
		Upper bound \cite{RSS}         & 6  & 18 & 30 & 40 & 50 \\ \hline
	\end{tabular}
	\caption{Equal overlap scenario simulated using algorithm in \cite{RSS} for $K=5$, the  number of common packets shared among two adjacent vehicle are $5$.}
	\label{AlgSimLB}
\end{table}

\begin{figure}
	\centering
	\includegraphics[scale=0.6]{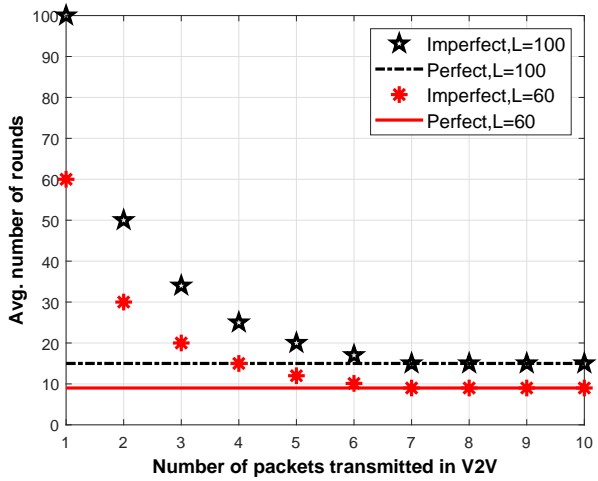}
	\caption{ Illustration of applying algorithm in \cite{RSS} in V2V for $l=2$, $K=5$ and $n=8.$}
	\label{NC_IC_Alg_lambda}
\end{figure}

\begin{figure}
	\centering
	\includegraphics[scale=0.6]{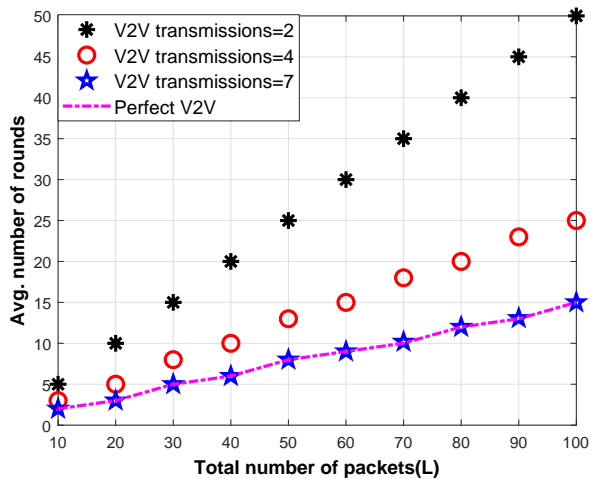}
	\caption{Illustration of applying algorithm in \cite{RSS} in V2V for $l=2$, $K=5$ and $n=8$ over various total number of packets.}
	\label{NC_IC_Alg_L}
\end{figure}

Furthermore, we examine the practicability of index codes in imperfect V2V  for different techniques of collaborative data download, namely feedback based technique and network coding based technique \cite{ZLS}.  
The decoding of index coded (IC) packets is not possible when the actual number of transmissions in the imperfect V2V phase is less than the minimum length of the index code.  So feedback based method of message dissemination in \cite{ZLS} fails in imperfect V2V conditions. However, in the network code (NC) based method, the decoding of IC packets at each node is not required since each IC packet itself is a valid NC packet. So  NC based method of message dissemination is suitable for IC under imperfect V2V conditions. Figure \ref{NC_IC_Alg_lambda}  and \ref{NC_IC_Alg_L} show the results obtained. Figure \ref{NC_IC_Alg_lambda} shows the average number of rounds that vehicles require to download a file with $L = 100$ and $60$ completely ($l = 2, K = 5$). Figure \ref{NC_IC_Alg_L} shows the average number of rounds required to download different file of size L when the number of exchanged packets in the V2V phase is constant.
%%%%%%%%%%%%%%%%
\subsection{Comparison with the scheme in \cite{ZLS} }
\begin{figure}
	\centering
	\includegraphics[scale=0.6]{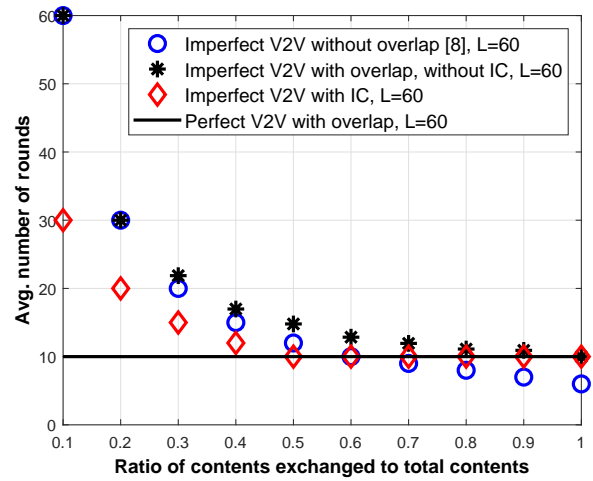}
	\caption{Illustration of advantage of IC in V2V for $l=2$, $K=5$ and $n=6$}
	\label{comparison}
\end{figure}

In this subsection we compare the results of collaborative message dissemination scenario under consideration with and without index coding. For a fair comparison we simulated the imperfect V2V set-up  in \cite{ZLS} with $K=5$, $l=2$ and $L=60$ by assuming the case that each vehicle receives $l$ independent packets in the R2V phase. Hence, for perfect V2V to be achieved, $K\times l=10$ independent packets are exchanged among the vehicles. This situation is marked as imperfect V2V without overlap in Figure \ref{comparison}. But in practice some vehicles in the cluster will be close to each other so that the adjacent vehicles may receive some common packets. For further simulations, the number of independent packets delivered by the RSU in each round is taken as $n=6$. Hence to achieve perfect V2V only $6$ independent packets need to be exchanged. Thus, more number of rounds are required for the complete delivery of the file. This situation is simulated without index coding and is labeled as imperfect V2V with overlap and without index coding in Figure \ref{comparison}. These common packets received could be considered as the side information possessed by each vehicle. Finally this case with overlap is simulated with index coding and it is clear from the simulations that less number of transmissions are required for achieving perfect V2V when index coding is used.

%\begin{figure}
%	\centering
%	\includegraphics[scale=0.6]{compare_8_pt4to1ER.eps}
%	\caption{Illustration of advantage of IC in V2V for $l=2$, $K=5$ and $n=6$}
%	\label{comparison2}
%\end{figure}
%%%%%%%%%%%%%%%%
%%%%%%%%%%%%%%%%%
\subsection{Simulation Using an Index Coding Example}
\begin{figure}
	\centering
	\includegraphics[scale=0.6]{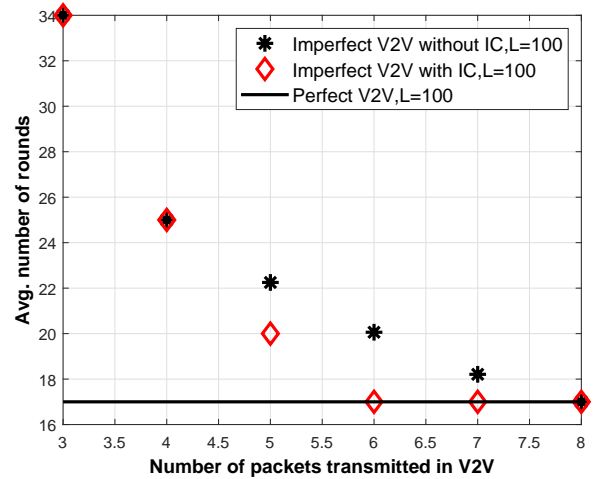}
	\caption{ Illustration of applying index coding in V2V with the example of $l=4$, $K=4$ and $n=8$}
	\label{EmplL_100}
\end{figure}

\begin{figure} 
	\centering
	\includegraphics[scale=0.6]{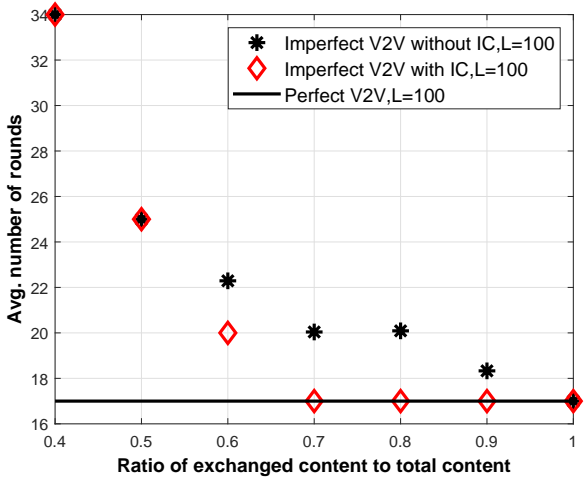}
	\caption{ Illustration of applying index coding in V2V with the example of $l=4$, $K=4$ and $n=8$ for different ratios of contents exchanged to the total contents.}
	\label{EmplER_100}
\end{figure}

Here  simulation is done with the example of index coding given in Section \ref{sec:motivating_example}. Imperfect V2V is assumed along with network coding (NC) at RSU \cite{ZLS}. The number of vehicles, $K=4$, the download capability $l=4$ and the total number of packets to be transferred, $n=8$. Index coding reduces transmissions to $6$ in this example, where $8$ transmissions were required without coding. Figure  \ref{EmplL_100} and \ref{EmplER_100} show the results obtained. From the results presented in figure  \ref{EmplL_100},  we get the total number of rounds needed for full file delivery to achieve perfect V2V condition. Figure  \ref{EmplER_100} presents the equivalent results redrawn for different ratios of contents exchanged to the total contents. 

%%%%%%%%%%%%%%%%
\subsection{Simulation using the optimal index coding matrix}
 In this section, we consider the equal overlap scenario with all the vehicles having the same download capability and the same number of common packets received by any two adjacent vehicles. The proposed lower bound for the typical scenario is compared with given bounds in \cite{RSS}. Also, the number of transmissions required using the algorithm in \cite{RSS} and the proposed encoding matrix is compared. The Tables \ref{AlgSimL_K5i5}, \ref{AlgSimL_K5i7} and \ref{AlgSimL_K7i7} show results for different values of $K$, $l$ and $i$. Recall that, $i$ is the number of common packets shared between two adjacent vehicles. It is evident from the simulation results that the proposed encoding scheme achieves the lower bound. 
\begin{table}[]
	\centering
	\begin{tabular}{|l|l|l|l|l|l|}
		\hline
		Total number of packets, $n$          & 10 & 20 & 30 & 40 & 50 \\ \hline
		Download capability, $l $             & 6  & 8  & 10 & 12 & 14 \\ \hline
		Lower bound \cite{RSS}                    & 5  & 13 & 21 & 29 & 37 \\ \hline
		Upper bound \cite{RSS}                    & 6  & 18 & 30 & 40 & 50 \\ \hline
		No.of transmissions using Alg. \cite{RSS}.   & 5  & 15 & 25 & 36 & 46 \\ \hline
		Lower bound $(n-i) $               & 5  & 15 & 25 & 35 & 45 \\ \hline
		No. of transmissions using the matrix $\textbf{L}$ & 5  & 15 & 25 & 35 & 45 \\ \hline
	\end{tabular}
	\caption{Equal overlap scenario simulated using algorithm in \cite{RSS} for $K=5$, the number of common packets shared among two adjacent vehicles are $5$.}
	\label{AlgSimL_K5i5}
\end{table}

\begin{table}[]
	\centering
	\begin{tabular}{|l|l|l|l|l|l|}
		\hline
		Total number of packets, $n$          & 12 & 22 & 32 & 42 & 52 \\ \hline
		Download capability, $l$              & 8  & 10 & 12 & 14 & 16 \\ \hline
		Lower bound \cite{RSS}                    & 5  & 13 & 21 & 29 & 37 \\ \hline
		Upper bound \cite{RSS}                    & 6  & 18 & 30 & 42 & 52 \\ \hline
		No. of transmissions using Alg. \cite{RSS} .   & 5  & 15 & 25 & 35 & 47 \\ \hline
		Lower bound $(n-i)$                & 5  & 15 & 25 & 35 & 45 \\ \hline
		No. of transmissions using the matrix $\textbf{L}$ & 5  & 15 & 25 & 35 & 45 \\ \hline
	\end{tabular}
	\caption{Equal overlap scenario simulated using algorithm in \cite{RSS} for $K=5$, the number of common packets shared among two adjacent vehicles are $7$.}
	\label{AlgSimL_K5i7}
\end{table}
\begin{table}[]
	\centering
	\begin{tabular}{|l|l|l|l|l|l|}
		\hline
		Total number of packets, $n$          & 14 & 28 & 42 & 56 & 70 \\ \hline
		Download capability, $l$              & 8  & 10 & 12 & 14 & 16 \\ \hline
		Lower bound \cite{RSS}                    & 7  & 19 & 31 & 43 & 55 \\ \hline
		Upper bound \cite{RSS}                    & 9  & 27 & 42 & 56 & 70 \\ \hline
		No. of transmissions using Alg. \cite{RSS}   & 7  & 21 & 36 & 50 & 65 \\ \hline
		Lower bound $(n-i)$                & 7  & 21 & 35 & 49 & 63 \\ \hline
		No. of transmissions using the matrix $\textbf{L}$ & 7  & 21 & 35 & 49 & 63 \\ \hline
	\end{tabular}
	\caption{Equal overlap scenario simulated using algorithm in \cite{RSS} for $K=7$, the number of common packets shared among two adjacent vehicles are $7$.}
	\label{AlgSimL_K7i7}
\end{table}

\begin{figure}[t]
	\centering
	\includegraphics[scale=0.6]{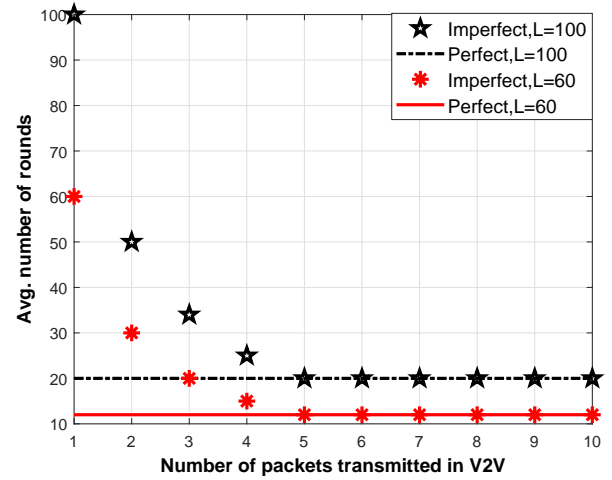}
	\caption{ Plot with encoding matrix in equal overlap scenario, $K=5$, $l=2$ and $i=1$ and $n=6$.}
	\label{NC_IClamda}
\end{figure}

\begin{figure}[t]
	\centering
	\includegraphics[scale=0.6]{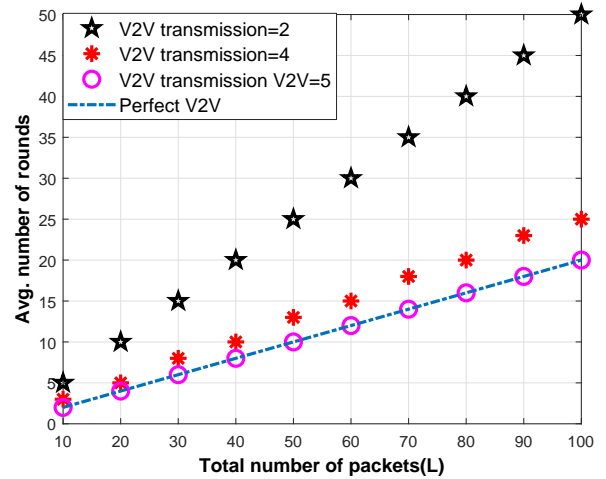}
	\caption{ Plot with encoding matrix in equal overlap scenario, $K=5$, $l=2$ and $i=1$ and $n=6$ for different file size.}
	\label{NC_IC_L}
\end{figure}

Simulation is performed for the index coding in V2V phase of the collaborative message dissemination problem with the designed encoding matrix. Imperfect V2V is assumed along with network coding (NC) at RSU. The simulation parameters are $K=5$, $l=2$ and $i=1$. Then total packets to be exchanged are $n=6$. Figure  \ref{NC_IClamda} and Figure \ref{NC_IC_L} show the results obtained.   Figure \ref{NC_IClamda} shows the average number of rounds required by vehicles to download a file of $L = 100$ and $60$. Figure \ref{NC_IC_L} shows the average number of rounds required to download different file of size $L$ when there is a constant number of packets exchanged in the V2V phase.

 For the design of $\textbf{L}$ matrix, each vehicle need to know the want set of other vehicles as well. So each vehicle has to announce the indices of packet received by them in R2V phase. Bandwidth requirements on transmitting these indices are too small when compared to the size of the packet. The size of each index is in one or two bytes where as the size of packet is in kB or MB. So, in the above simulations, we don't consider the bandwidth required for the feedback.
%%%%%%%%%%
\section{Conclusion}
In this paper, we studied the problem of index coding in V2V phase of collaborative message dissemination protocol.  We obtained the lower bound on the number of transmissions required in the V2V communication phase of the above problem when the download capability of the vehicles are the same in the R2V phase. An optimal index code for a typical VANET scenario is also proposed. When the links are error-prone, we constructed an optimal linear error correcting index code for this scenario. As future work, this coding can be extended to more general scenarios. It is also shown that, when associated with index codes, the network-based dissemination technique has more practicality over feedback-based schemes in imperfect V2V.  

%%%%%%%%%%%%%%%%%%%%%%%%%%%%%%%%%%%%%%%%%%%%%%%%%%%%%%%%%%%%%%%%%%%%%
\section*{Acknowledgment}
This work was supported partly by the Science and Engineering Research Board (SERB) of Department of Science and Technology (DST), Government of India, through J.C. Bose National Fellowship to B. Sundar Rajan.

%%%%%%%%%%%%%%%%%%%%%%%%%%%%%%%%%%%%%%%%%%%%%%%%%%%%%%%%%%%%%%%%%%%
\begin{IEEEbiography}[{\includegraphics[width=1in,height=1.25in,clip,keepaspectratio]{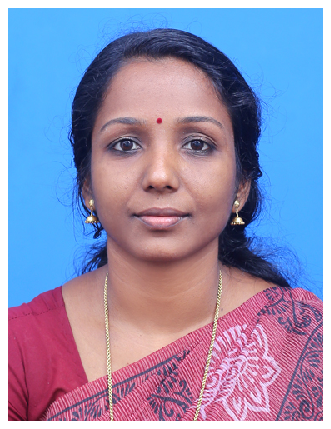}}]{Jesy Pachat} was born in Kerala, India. She received B.Tech degree in Electronics and Communication Engineering in 2004 from Mahatma Gandhi University, Kerala, and M.Tech Degree in Telecommunication from National Institute of Technology, Calicut in 2014.  She has been working as faculty in Govt. college of engineering Kannur since 2010. She is currently pursuing her Ph.D. degree in the Department of Electronics and Communication Engineering, National Institute of Technology, Calicut. Her primary research interests include network coding, index coding, and coding theory.
\end{IEEEbiography}

\begin{IEEEbiography}[{\includegraphics[width=1in,height=1.25in,clip,keepaspectratio]{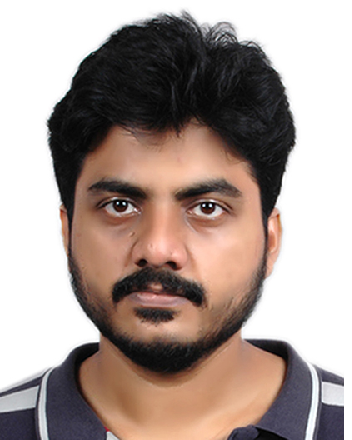}}]{Nujoom Sageer Karat} was born in Kerala, India. He received his B.Tech degree in Electronics and Communication Engineering in 2012 from National Institute of Technology, Calicut, and Ph.D degree from the Department of Electrical Communication Engineering, Indian Institute of Science. He worked as an Engineer in Renesas Mobile India Private Limited. His primary research interests include index coding, coded caching and error correcting index coding.  
\end{IEEEbiography} 

\begin{IEEEbiography}[{\includegraphics[width=1in,height=1.25in,clip,keepaspectratio]{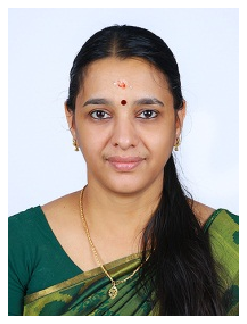}}]{Deepthi P.P.} was born in Kerala, India. She received M. Tech Degree in Instrumentation from Indian Institute of Science, Bangalore in 1997 and Ph.D. from National Institute of Technology Calicut in 2009 in the field of Secure Communication. She is currently working as Professor in the Department of Electronics and Communication Engineering, National Institute of Technology Calicut. Her current interests include Cryptography, Information Theory, Coding Theory, and Multimedia Security and Secure Signal Processing. 
\end{IEEEbiography} 

\begin{IEEEbiography}[{\includegraphics[width=1in,height=1.25in,clip,keepaspectratio]{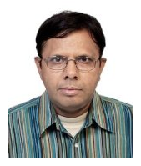}}]{B. Sundar Rajan} (S'84-M'91-SM'98-F'14) was born in Tamil Nadu, India. He received the B.Sc. degree in mathematics from the Madras University, Madras, India, in 1979, the B.Tech. degree in electronics from the Madras Institute of Technology, Madras, in 1982, and the M.Tech. and Ph.D. degrees in electrical engineering from the IIT Kanpur, Kanpur, in 1984 and 1989, respectively. He was a faculty member at the Department of Electrical Engineering, IIT Delhi, New Delhi, from 1990 to 1997. He has been a Professor with the Department of Electrical Communication Engineering, Indian Institute of Science, Bangalore, since 1998. His primary research interests include space-time coding for MIMO channels, distributed space-time coding and cooperative communication, coding for multiple-access and relay channels, and network coding.

Dr.~Rajan was an Editor of the \textsc{IEEE Wireless Communications Letters} (2012-2015), an Associate Editor of Coding Theory for the \textsc{IEEE Transactions On Information Theory} (2008-2011 and 2013-2015), and an Editor of the \textsc{IEEE Transactions on Wireless Communications} (2007-2011). He served as a Technical Program Co-Chair of the IEEE Information Theory Workshop (ITW’02), held in Bangalore, in 2002. He is a fellow of IEEE, the Indian National Academy of Engineering, the Indian National Science Academy, the Indian Academy of Sciences, and the National Academy of Sciences, India. He is a recipient of Prof. Rustum Choksi Award by I.I.Sc for Excellence in Research in Engineering for the year 2009, the IETE Pune Center’s S.V.C Aiya Award for Telecom Education in 2004, and Best Academic Paper Award at the IEEE WCNC 2011. He is a J.C. Bose National Fellow (2016-2020) and a member of the American Mathematical Society.
\end{IEEEbiography}

\begin{thebibliography}{1}
\bibitem{sur}
Z. MacHardy, A. Khan, K. Obana and S. Iwashina, ``V2X Access Technologies: Regulation, Research, and Remaining Challenges," in \emph{IEEE Communications Surveys \& Tutorials}, vol. 20, no. 3, pp. 1858-1877, thirdquarter 2018.
	
\bibitem{3gppRel14}
3GPP, ``Architecture enhancements for V2X services (v14.3.0, Release14)," 3GPP, Tech. Rep. 23.285, June 2017.	

\bibitem{v2xAna}
M. Gonzalez-Martín, M. Sepulcre, R. Molina-Masegosa and J. Gozalvez, ``Analytical Models of the Performance of C-V2X Mode 4 Vehicular Communications," in \emph{IEEE Transactions on Vehicular Technology}, vol. 68, no. 2, pp. 1155-1166, Feb. 2019.


\bibitem{veh_mag}
R. Molina-Masegosa and J. Gozalvez, ``LTE-V for Sidelink 5G V2X Vehicular Communications: A New 5G Technology for Short-Range Vehicle-to-Everything Communications," in \emph{IEEE Vehicular Technology Magazine}, vol. 12, no. 4, pp. 30-39, Dec. 2017.


\bibitem{coexist}
P. Wang, B. Di, H. Zhang, K. Bian and L. Song, ``Cellular V2X Communications in Unlicensed Spectrum: Harmonious Coexistence With VANET in 5G Systems," in \emph{IEEE Transactions on Wireless Communications}, vol. 17, no. 8, pp. 5212-5224, Aug. 2018.


\bibitem{dsrc}
J. B. Kenney, ``Dedicated Short-Range Communications (DSRC) Standards in the United States," in \emph{Proceedings of the IEEE}, vol. 99, no. 7, pp. 1162-1182, July 2011.


\bibitem{NC}
R. Ahlswede, Ning Cai, S. -. R. Li and R. W. Yeung, ``Network information flow," in \emph{IEEE Transactions on Information Theory}, vol. 46, no. 4, pp. 1204-1216, July 2000.

\bibitem{ZLS}
W. Zhu, D. Li and W. Saad, ``Multiple Vehicles Collaborative Data Download Protocol via Network Coding," in \emph{IEEE Transactions on Vehicular Technology}, vol. 64, no. 4, pp. 1607-1619, April 2015.

 \bibitem{col1}
 M. H. Firooz and S. Roy, ``Collaborative downloading in VANET using Network Coding," \emph{2012 IEEE International Conference on Communications (ICC)}, Ottawa, ON, 2012, pp. 4584-4588.

 \bibitem{col2}
 F. Ye, S. Roy and H. Wang, ``Efficient Data Dissemination in Vehicular Ad Hoc Networks," in \emph{IEEE Journal on Selected Areas in Communications}, vol. 30, no. 4, pp. 769-779, May 2012.
 
 \bibitem{col3}
 A. Nandan, S. Das, G. Pau, M. Gerla and M. Y. Sanadidi, ``Co-operative downloading in vehicular ad-hoc wireless networks," \emph{Second Annual Conference on Wireless On-demand Network Systems and Services}, St. Moritz, Switzerland, 2005, pp. 32-41.
  
 \bibitem{col4}
 S. Ahmed and S. Kanhere, ``VANETCODE: Network coding to enhance cooperative downloading in vehicular Ad-Hoc networks,” in\textit{ Proc. IWCMC}, Vancouver, BC, Canada, Jul. 2006, pp. 527–532.

\bibitem{GXGC}
Y. Gao, X. Xu, Y. L. Guan and P. H. J. Chong, ``V2X Content Distribution Based on Batched Network Coding With Distributed Scheduling," in \emph{IEEE Access}, vol. 6, pp. 59449-59461, 2018.       
       

\bibitem{RSS}
S. E. Rouayheb, A. Sprintson and P. Sadeghi, ``On coding for cooperative data exchange," \emph{2010 IEEE Information Theory Workshop on Information Theory (ITW 2010, Cairo)}, Cairo, 2010, pp. 1-5.

\bibitem{D2D1}
Y.~D.~Lin and Y.~C.~Hsu, ``Multihop cellular: A new architecture for wireless communications,'' in \emph{Proc. IEEE INFOCOM 2000, Conference on Computer Communications. Nineteenth Annual Joint Conference of the IEEE Computer and Communications Societies}, Tel Aviv, Israel, vol. 3, pp. 1273-1282, April 2000.

\bibitem{JCM}
M.~Ji, G.~Caire and A.~F.~Molisch, ``Fundamental Limits of Caching in Wireless D2D Networks," in \emph{IEEE Transactions on Information Theory}, vol. 62, no.2, pp. 849-869, Feb 2016.

\bibitem{MPRGR}
N. Milosavljevic, S. Pawar, S. E. Rouayheb, M. Gastpar, and K. Ramchandran, ``Efficient algorithms for the data exchange problem," in \emph{IEEE Transactions on Information Theory}, vol. 62, no. 4, pp. 1878 – 1896, Apr. 2016.

 \bibitem{BiK}
 	Y. Birk and T. Kol, ``Coding-on-demand by an informed source (ISCOD) for efficient broadcast of different supplemental data to caching clients," \textit{IEEE Trans. Inf. Theory}, vol. 52, no. 6, pp. 2825-2830, June 2006.
 	
	\bibitem{DSC}
	Son Hoang Dau, V. Skachek, and Yeow Meng Chee, ``Error correction for index coding with side information," {\textit{IEEE Trans. Inf. Theory}}, vol. 59, no. 3, pp. 1517-1531, 2013. 
	
	\bibitem{KTR}
N. S. Karat, A. Thomas and B. S. Rajan, ``Error Correction in Coded Caching With Symmetric Batch Prefetching," in \emph{IEEE Transactions on Communications}, vol. 67, no. 8, pp. 5264-5274, Aug. 2019.

\bibitem{KSR}
N. S. Karat, S. Samuel and B. S. Rajan, ``Optimal Error Correcting Index Codes for Some Generalized Index Coding Problems," in \emph{IEEE Transactions on Communications}, vol. 67, no. 2, pp. 929-942, Feb. 2019.

\bibitem{KTR2}
N. S. Karat, A. Thomas and B. S. Rajan, ``Optimal Error Correcting Delivery Scheme for an Optimal Coded Caching Scheme with Small Buffers," \emph{2018 IEEE International Symposium on Information Theory (ISIT)}, Vail, CO, 2018, pp. 1710-1714.

\bibitem{KDTR}
N. S. Karat, S. Dey, A. Thomas and B. S. Rajan, ``An Optimal Linear Error Correcting Delivery Scheme for Coded Caching with Shared Caches," \emph{2019 IEEE International Symposium on Information Theory (ISIT)}, Paris, France, 2019, pp. 1217-1221.

\bibitem{Gra}
M.~Grassl, ``{Bounds on the minimum distance of linear codes and quantum codes},'' Online available at {http://www.codetables.de}, 2007.
 	
\end{thebibliography}
\end{document}